\documentclass{aa}
\usepackage{txfonts}
\usepackage[]{graphicx}
\usepackage{subcaption}
\usepackage{float}
\usepackage{comment}

\bibpunct{(}{)}{;}{a}{}{,}

\begin{document}
\title{Determination of the position angle of stellar spin axes}
\subtitle{}
\author{A-L Lesage\inst{1,2} \and  G. Wiedemann\inst{1} }
\institute{Hamburger Sternwarte,
              Gojenbergsweg 112, 21029 Hamburg, Germany
              \and Leiden Observatory, Leiden University, Niels Bohrweg 2, 2333CA Leiden, The Netherlands \\ \email{lesage@strw.leidenuniv.nl} }
\date{}
\abstract
{Measuring the stellar position angle provides valuable information on binary stellar formation or stellar spin axis evolution.
}
{ We aim to develop a method for determining the absolute stellar position angle using spectro-astrometric analysis of high resolution long-slit spectra. The method has been designed in particular for slowly rotating stars. We investigate its applicability to existing dispersive long-slit spectrographs, identified here by their plate scale, and the size of the resulting stellar sample.
}
{ The stellar rotation induces a tilt in the stellar lines whose angle depends on the stellar position angle and the orientation of the slit. We developed a rotation model to calculate and reproduce the effects of stellar rotation on unreduced high resolution stellar spectra. Then we retrieved the tilt amplitude using a spectro-astrometric extraction of the position of the photocentre of the spectrum. Finally we present two methods for analysing the position spectrum using either direct measurement of the tilt or a cross-correlation analysis. }
{ For stars with large apparent diameter and using a spectrograph with a small plate scale, we show that it is possible to determine the stellar position angle  directly within $10^\circ$ with a signal-to-noise ratio of the order of 6. Under less favourable conditions, i.e. larger plate scale or smaller stellar diameter, the cross-correlation method yields comparable results.}
{ We show that with the currently existing instruments, it is possible to determine the stellar position angle of at least 50 stars precisely, mostly K-type giants with apparent diameter down to 5 milliarcseconds.
If we consider errors of around $10^\circ$ still acceptable, we may include stars with apparent diameter down to 2 mas in the sample that then comprises also some main sequence stars.} 

\maketitle

\section{Introduction}\label{sec:Intro}

Angular momentum vectors are usually assumed to be randomly oriented during the star formation process leading to a uniform distribution of stellar spin axis orientations.  
However, there might be a physical process -- for instance magnetic fields (\cite{Hennebelle})-- that would lead to a preferred orientation on the scale of a star forming region despite chaotic stellar formation (\cite{Bate}). 
As a result, measuring the absolute orientation of the stellar spin axis would provide valuable information for constraining the models for stellar formation and stellar evolution, such as spin axis distribution inside a stellar cluster, spin axes alignment of binaries, and evolution of the stellar spin axis orientation with the age of the star. 


The spin axis orientation is defined by two angles: inclination $i$, which is the angle formed between the spin axis and the line of sight with the observer, and the position angle, which is the projection of the spin axis on the sky and measured from north to east. 
In this paper we concentrate on determining the position angle.

For eclipsing objects, such as binary stars or a star with an exo\-planet, measuring the Rossiter-McLaughlin effect, i.e. the dimming of light as a function of wavelength, provides additional information on the system geometry. 
During the eclipse,  a portion of the rotating star is blocked by its companion, causing a weakening in the corresponding Doppler component of the stellar absorption lines. The shape of this spectral distortion depends on the angle between the stellar spin axes, in the case of binaries, or stellar and orbital spin axes, in the case of exoplanets.
 
This effect is currently exploited in many cases of transiting exoplanets to deliver the spin-orbit alignment. However, the associated error is quite important. To date, the mean error to spin-orbit measurements is close to $10^\circ$. There are no constraints available for the orientation of the absolute stellar spin axis.

The technical improvements in long baseline interferometry during the past decade have opened the possibility of resolving and imaging stellar surfaces quite accurately. Several teams have observed the absolute spin axis orientation of a dozen early-type stars: Altair (\citet{Monnier}), Vega (\citet{Peterson}), and Achenar (\citet{deSouza}). 
The spin axis orientation is actually a by-product of determining the gravitational limb darkening coefficient for these fast rotating stars. 
Gravitational limb darkening causes a shift of the photocentre towards the poles. 
The amplitude of the  shift is highly dependent on the rotational velocity of the star. 
For slowly rotating objects, which are mainly late-type stars, the method soon becomes unreliable because the shift is no longer measurable.  

\citet{Petrov} devised an alternative interfero\-metric method of determining the position angle on large,  slowly rotating stars using differential interferometry. The method combines speckle interferometry and spectroscopy. 
They show that using a high-to-moderate-resolution spectrograph, coupled with an interferometer, the position angle could be measured within $10^\circ$ precision. 
The method was tested successfully on Aldebaran (\citet{Lagarde}). However, owing to the short speckle integration time, its use remained limited to the brightest stars (Lagarde, private communication). 

There is currently a lack of observational methods for determining stellar spin orientation for slowly rotating stars. 
We investigate the potential of spectro-astrometry to measure the position angle of these stars, with focus on late-type giants. 
We aim here to define the optimal conditions, instrumental and observational, in order to accurately measure, within $10^\circ$ errors, the position angle of a large sample of stars and the required signal-to-noise ratio (S/N).

Spectro-astrometry concentrates on the spatial information contained in the spectrum by measuring the wavelength dependent position of the photocentre of an object. 
It is a powerful tool that allows us to reach sub-diffraction resolution using standard directly-fed spectrographs from otherwise unresolved sources.  
In addition, it is an instrumentally easy method, since it only requires a stable spectrograph and a detector. 
However, it also requires a special observing practice, since different slit orientations are required for the analysis.
Nonetheless it can be applied to a wide variety of science cases. 
It has been used successfully in the study of unresolved binaries (\citet{Bailey, Baines}), of outflows from young objects down to the AU scale (\cite{Takami}), and of jets from brown dwarfs (\citet{Whelan}). Furthermore, \citet{Voigt} applied the method for mapping stellar spots on red giant surfaces using \textsc{Crires}\footnote{Cryogenic high Resolution Infrared \'Echelle Spectrograph}.

In a long-slit spectrograph, the spatial information is conserved from the slit to the detector. Indeed the spatial scale on the detector can be directly derived from the optics of the spectrograph. 
The position of the photocentre is determined either by fitting a 1D Gaussian over the order or by a weighted arithmetic mean for each wavelength. 
As a result, even if the source is unresolved, as long as it presents an asymmetric spectral energy distribution, the position of the photocentre will diverge from the continuum at certain wavelengths. 
Furthermore, it can be measured to sub-pixel precision.  
The only limitation is set by the pixels inhomogeneities of the detector and the signal-to-noise ratio. 
The precision reached in the position of the photocentre is approximated by \begin{math} \sigma = 0.5 \times \text{FWHM}\times \text{S/N}^{-1} \end{math}.

For instance, with an average seeing of 1", assuming a detector with an one-to-one conversion from photon to electron, a single exposure can achieve up to 40 000 photons counts in the spectrum. Then using the previous formula, the precision in the position is estimated close to 2.5 milliarcseconds (mas). This scale is comparable with the apparent diameter of our closest or largest neighbour stars.

\vspace{1em}
This paper is organised as follows. In Sect \ref{sec:2ddoppler}, we define a 2D model for  stellar rotation adapted for use in the spectro-astrometric analysis. In Sect 3, we describe the simulation set-up for the construction of the 2D spectrum and the extraction of the spectro-astrometric signal. Section 4 presents the results of our simulations, which are discussed in Sect 5. 

\section{Two-dimensional stellar rotation model}\label{sec:2ddoppler}

The 2D Doppler model for stellar rotation is a tool for converting the star from the slit coordinates ($X_s$, $Y_s$) into the detector coordinates ($\lambda$, $Y_d$). It is based on the demonstration of rotational line broadening by \citet{Gray} except that we take the spatial size of the star and the orientation of the stellar spin axis with respect to the slit into account. 

\begin{figure}
\centering
\includegraphics[width = 0.35 \textwidth]{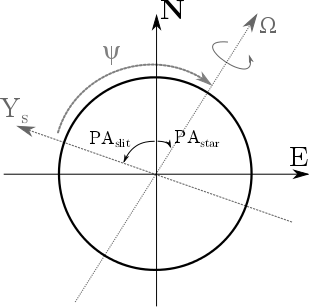}
\caption[Notation convention]{Notation conventions adopted in this paper showing the cross-section of the star projected on the sky. The absolute stellar position angle $PA_{star}$ is defined from north to east. The slit is represented by the spatial axis $Y_S$, which may not be aligned with  the northern axis by an angle $PA_{slit}$. $\psi$ is the angle formed between the slit axis and the stellar position angle.  }
\label{fig:convention}
\end{figure}

We made the following assumptions. The projected position angle of the slit on the sky is known at any time by the observer. The star rotates as a rigid body around its spin axis with an angular velocity $\Omega$, counted positively counter-clockwise. 
In addition, we define $\psi$ as the projected angle on the sky between the stellar spin axis and the slit spatial axis $Y_s$, counted as positive from north to east. The notations are defined in Fig~\ref{fig:convention}. Finally, each point of the star is described by the coordinates X and Y linked to the stellar diameter by the straightforward relation: \begin{math}X^2 + Y^2 = R_{star}^2 \end{math} where $R_{star}$ is the angular radius of the star.

The wavelength shift caused by the stellar rotation is calculated for each point of the star projected on the slit and includes the rotation of the reference frames of the star and the slit:

\begin{align}\label{eq:dwdop}
 \delta \lambda_D (X,Y, \psi) & = \frac{\Omega \sin i \,\lambda}{c} \left ( X \cos \psi - Y \sin \psi \right ).
\end{align}

Since spectro-astrometry allows us to reach the milliarcsecond scales, the star can no longer be considered as a point source on the slit even if it is not resolved. 
Its spatial extension is preserved along the spatial axis.
However, a broadening along the spectral axis is translated into a broadening of the absorption lines. 
Indeed variations in the position $X_s$ cause changes in the incident angle $\alpha$ of the grating, which are in turn translated into  wavelength shifts. 
Using the \'Etendue relation between star and grating, \begin{math}  D \phi_{star} = L \cos \alpha \, \delta \alpha \end{math}, where D is the diameter of the telescope and L the effective grating length,  and inserting it into the differential of the grating equation \begin{math} m\, \delta \lambda = d \cos \alpha \; \delta \alpha \end{math}, where m is the spectral order and d the grating constant, we obtain the geometrical shift as a function of the parameters of the spectrograph:
\begin{equation}
\delta \lambda_{geo}(X) = \frac{d D \phi_{star}(X)}{m L}.
\end{equation}
The parameter $\phi_{star}$ is the apparent angular diameter, in arcseconds,  of the star on the slit without the seeing disk. This is noted directly X from now on. 
For nearby slowly rotating stars, with $V_{rot} < 20$ km/s, observed with a high resolution spectrograph, $R > 60000$, the geometrical shift is typically two orders of magnitude lower than the Doppler component. 

The total wavelength shift for each point on the star is then the sum of both contributions $\delta \lambda_D$ and $\delta \lambda_{geo}$:
\begin{align}\label{eq:totalshift}
\Delta \lambda (X, Y, \psi) &= \frac{d D X}{mL} + \frac{\Omega\, \sin i\; \lambda}{c}\times \left ( X \cos\psi - Y \sin \psi \right ) .
\end{align} 

The overall shape of the intensity distribution is evaluated here by a linear limb-darkening law:

\begin{equation}
I_c = I_o (1 - \epsilon + \epsilon \cos \mu)
\end{equation}
where $I_o$ is the intensity at the centre of the star, $\epsilon$ the limb-darkening coefficient and $\mu$ the angular limb distance defined as 
\begin{equation}\label{eq:cosmu}
 \cos \mu = \sqrt{R_{star}^2 - \left (X^2+ Y^2 \right)}/R_{star} .
\end{equation}

\begin{figure*}[ht]
\centering
\resizebox{\hsize}{!}{\includegraphics[width = 17cm]{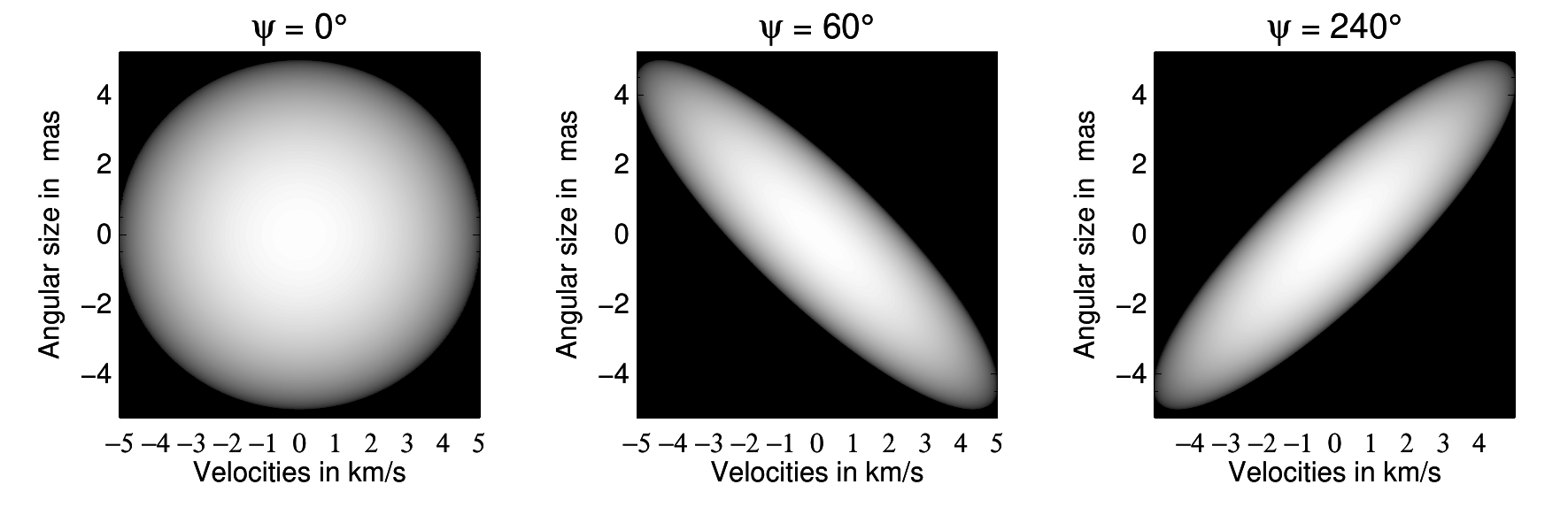}}
 \caption[Shape of the line broadening function in dependency of $\psi$]{Shape of the line broadening function in dependency of $\psi$. The model assumes a rotational velocity $V_{rot} \sin i =5 $km/s, an apparent radius $R_{star} = 5$ mas,  and a limb-darkening coefficient $\epsilon = 0.9$. The spectrograph parameters, used for the geometrical shift, are a telescope size of 1 m, an incoming angle $\alpha$ of $65^\circ$, a standard grating constant, and an observation in the 100\textsuperscript{th} order.}
\label{fig:doppler2d}
\end{figure*}

Substituting X from Equation (\ref{eq:totalshift})  into Equation (\ref{eq:cosmu}), the right term of the linear limb-darkening law is then
\begin{align}
D(\lambda, Y, \psi) & = \frac{I_c(\lambda, Y, \psi)}{I_o} \\ \nonumber
	& = 1 - \epsilon + \epsilon \frac{ \sqrt{R_{star}^2 - \left ( \left (\frac{\Delta \lambda (X,Y,\psi) - K\,Y}{K'} \right )^2 + Y^2 \right ) }}{R_{star}} 
\end{align}
where K and K' are two constant terms resulting from the reformulation of Equation (\ref{eq:totalshift}). 
The shape of the intensity distribution now depends on $\sin \psi$ as seen in Fig \ref{fig:doppler2d}. It is a slanted ellipse with an inclination angle $\xi$:
\begin{equation}\label{eq:tiltincrease}
 \tan \xi = \frac{V_{rot} \sin i \; \lambda \;\sin \psi}{c \, R_{star}}.
\end{equation}
To determine the stellar position angle, it is required that the dependency of $\xi$ on $\sin \psi$ be measured.  

If we collapse the profile along the spectral direction, we would retrieve the classical broadening profile used in traditional spectroscopy.

\section{Simulation set-up}

\begin{figure}
\resizebox{\hsize}{!}{\includegraphics{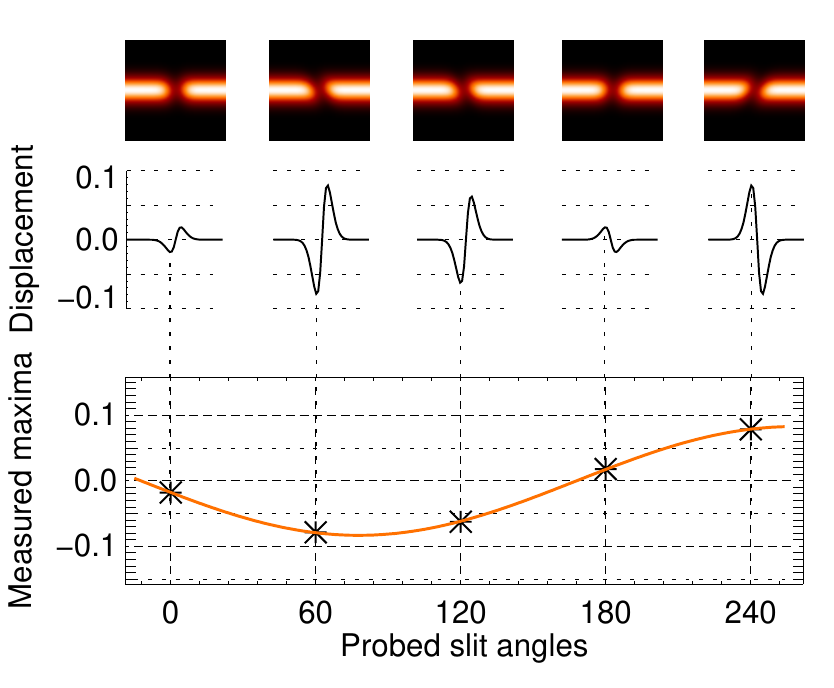}}
\caption{Step-by-step explanation of the position angle determination: spectra are recorded for different slit angles. Top: the 2D spectrum of one absorption line, tilted as a function of $\psi$. The tilts are exaggerated here in order to be visible. Middle: the position spectrum measured around the absorption line presenting the typical displacement of stellar rotation. Bottom: using the maxima of the displacement on the position spectrum, we fit a sine function whose phase is the stellar position angle. Here $\text{PA}_\text{star}$ was set at 12$^\circ$.  }
\label{fig:principle}
\end{figure}

\begin{figure}
\resizebox{\hsize}{!}{\includegraphics{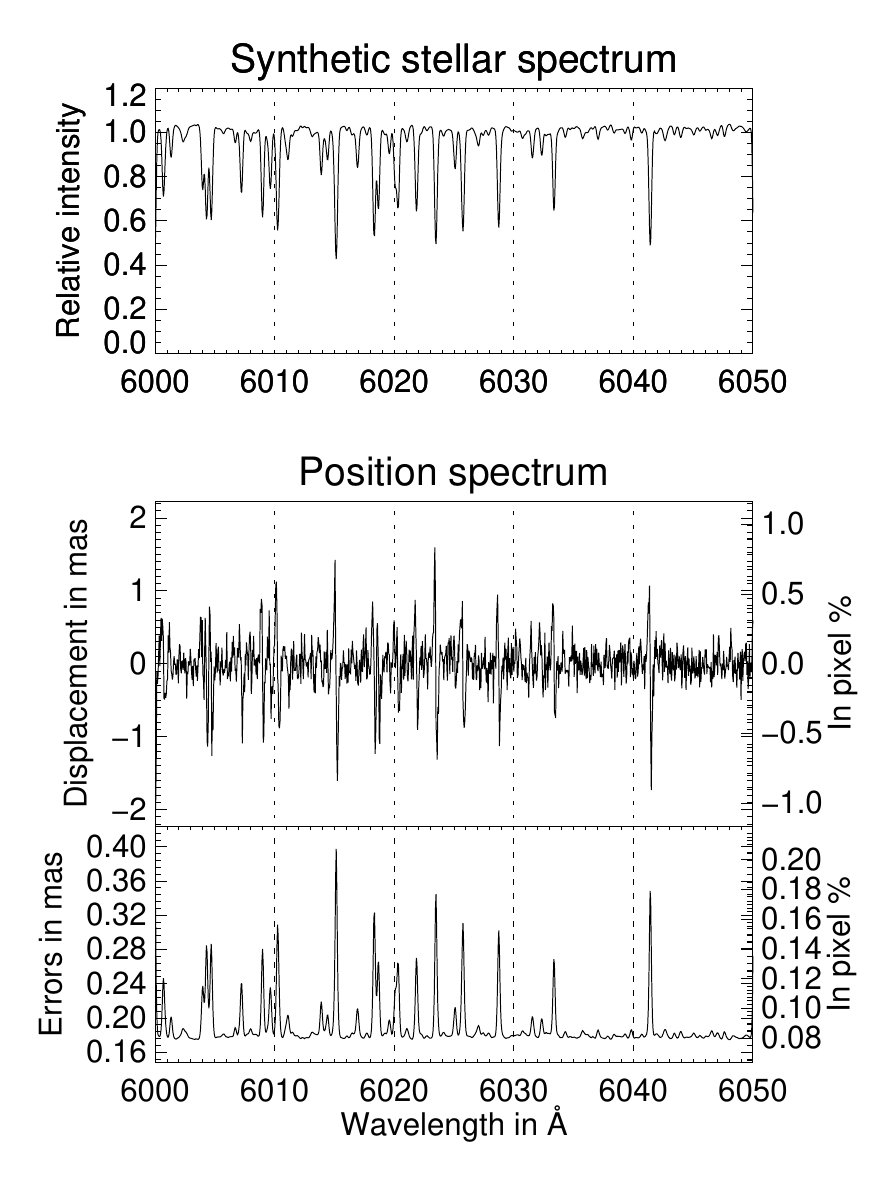}}
\caption[]{Top: the intensity spectrum extracted from the synthetic 2D spectrum of the cool giant for $\psi = 90^\circ$. Middle: the corresponding position spectrum. The pixel plate scale is $P_s = 0.2$"/pixel, the seeing is set at 1" ($S_{eeing}= 5\, P_s$) and the signal to noise ratio of the 2D spectrum is fixed to 150. In this example the apparent stellar diameter is set to be as large as Aldebaran's: 20 mas, which permits a visual identification of the displacement of the photocentre at the line position. Bottom: the errors on the position spectrum in mas. }
\label{fig:blo}
\end{figure}

An observed 2D spectrum is the convolution of the intrinsic stellar spectrum and the rotational model, and it is further affected by instrumental and seeing effects:
\begin{equation}
I_{obs} = \{ I_{int}(\lambda, Y) \otimes D(\lambda, Y, V_{rot}\sin i, \psi)\} \otimes S_{eeing}(\lambda, Y). \end{equation}
For simplicity we included here the instrumental profile to the atmospheric effects on the term $S_{eeing}$. 
The working principle of the method is described in Fig \ref{fig:principle}. 
The inclination of the rotational profile is transmitted to the stellar absorption lines in the observed spectrum: they are tilted by a small angle in dependency to $\sin \psi$. 
The tilt is expected to be in the sub-pixel range. 
However, it has a unique signature on the position spectrum since it causes an identical displacement for all the stellar lines (see Fig \ref{fig:blo}). 
The amplitude of the displacement of the photocentre is proportional to the line inclination. 
The absolute position angle of the star ($\text{PA}_{\text{star}}$) is retrieved by monitoring the variation in the line tilt that has the dependency of \begin{equation}
\sin \psi = \sin (\text{PA}_{\text{slit}} + \text{PA}_{\text{star}}).
\end{equation}

\subsection{Construction of the 2D spectrum}\label{sec:bla}

As an intrinsic 1D stellar spectrum, we took a synthetic \textsc{phoenix} spectrum (\cite{Hauschildt}) calculated for a cool gi\-ant like Aldebaran using the following parameters: an effective temperature of 4\,000K, a surface gravity \begin{math} \log(g) = 1.8 \end{math}, and solar-like metallicity. It was calculated under the assumption of thermal equilibrium in the outer layers of the stellar atmosphere. The synthetic spectra has an average resolution of 120\,000 over a wavelength range of 5 nm in the visible. The 1D spectrum is expanded homogeneously in the spatial direction by matrix multiplication with a Gaussian profile that corresponds to the apparent stellar diameter. 

The instrument is simulated using a standard R2 reflection grating for a cross-dispersion spectrograph and a spectral order of interest of m = 100. 
The remaining parameters describing the instrument, i.e. length, incident angle, and primary mirror size, are chosen in order to match the desired spectral resolution. 

The resolution of the spectrograph is set by  binning the synthetic data. Consequently, we explored two resolutions of 60\,000 and 120\,000 for a constant stellar rotation velocity of 5 km/s. 
In addition, we also varied the apparent stellar diameter from 2.5 mas to 30 mas (equivalent to the diameter of Betelgeuse),  and the plate scale, $P_s$, from 0.086"/pix as used in \textsc{Crires - Macao\footnote{Multi-Applications Curvature Adaptive Optics}} (\cite{Kaeufl}) to 0.5"/pix representing the Th\"uringer Landessternwarte Spektrograph (hereafter TL-Spectrograph). 
We assumed that the plate scale is generally defined according to the local seeing conditions on site, with a sampling of five pixels per stellar full width half maximum (FWHM). 
Finally the resulting spectrum is convolved again with a seeing function defined in the $(\lambda, Y_d)$ coordinates. 
The simulations were done under the assumption of optimal slit configuration; i.e., the slit width is always smaller than or equal to the seeing FWHM. Furthermore, the seeing does not bend the spectral lines. 
We discuss later how the results are affected in a real case and how to analyse them consequently.

Finally Gaussian noise is applied to the 2D spectrum. All simulations presented here are realized with a raw signal-to-noise ratio of 150 unless noted otherwise. The raw S/N is estimated as the photon noise over the spectral order.

\subsection{Signal extraction}

The position spectrum $B(\lambda)$ is retrieved by measuring the position of the photocentre using an  arithmetic weighted mean:
\begin{align*}
B (\lambda ) & = \frac{\sum_i y_i F_i}{\sum_i F_i} \\ 
\delta B(\lambda) & = \sqrt{\frac{\sum_i F_i (y_i - B(\lambda))^2)}{(\sum_i F_i)^2}}.
\end{align*}
The intensity spectrum is obtained by collapsing the 2D spectrum along the spatial direction.  
We studied two different me\-thods of retrieving the amplitude variations of the signal from the position spectrum. 

In the first method, we directly measured the amplitude variations of the position of the photocentre for one deep and narrow absorption line. 
The errors in the amplitude are derived directly from the seeing FWHM and the S/N of the spectrum. However, this method can only be applied if the signal amplitude is signi\-ficantly higher than the scatter of the continuum. 

The second method uses a cross-correlation analysis to extract the signal amplitude and to monitor its variations.
The noiseless position spectrum can be derived from the intensity spectrum $I(\lambda)$ via the contrast of the spectrum \begin{math} |(\mathrm{d}I\,(\lambda, V_{rot}\sin i)/\mathrm{d}\lambda)^2| \end{math}:
\begin{equation}\label{eq:contrast}
B(\lambda, \psi) = \frac{\mathrm{d}I \,(\lambda, V_{rot}\sin i)}{\mathrm{d}\lambda} \times F(R_\text{star}, S_{eeing}) \times \sin \psi
\end{equation}
where F is dependent on both $S_{eeing}$ the average seeing FWHM during the observation run and the apparent size of the star. It is assumed to be constant during the observation for a given orientation. The interpretation of Equation (\ref{eq:contrast}) implies that only deep thin lines, which are characterized by a high contrast value, will maximize the displacement on the position spectrum. 

Therefore, we calculated the cross-correlation function between the position spectrum $B(\lambda, \psi)$ and the derivative of the intensity spectrum \begin{math}\mathrm{d}I(\lambda)/\mathrm{d} \lambda \end{math} to retrieve the amplitude variations. 
For each slit orientation, we extracted the maximum of the cross-correlation function $C_{xy}$, which should be located at the centre of the line. 
The errors of the cross-correlation function were calculated using \begin{math} \sigma = (1- C_{xy})^2 / (2 W) \end{math} where W is the average line width in pixel. In the next section, we discuss the validity of this formula for the error estimations.

Finally we applied a Levenberg-Marquardt algorithm (\textsc{mpfit}) from \cite{Markwardt} on the selected points to perform the least-square fit of the sine function as illustrated in Fig \ref{fig:principle}. 
The retrieved phase of the fit yields the value of the absolute stellar position angle.

\section{Results}\label{sec:result}

\subsection{Expectations}
The measured displacement on the position spectrum is always smaller than the apparent stellar radius. 
Figure \ref{fig:blo} illustrates this phenomenon. With the given pixel plate scale, $P_s$, of 0.2"/pixel, and a stellar radius of 10 mas, one could expect a maximum displacement of the photocentre of \begin{math} \delta = R_\text{star}/P_s = 0.05 \; \text{pixel} \end{math} to occur when slit and spin axis are orthogonal.
However, the measured displacement is always more than two times smaller. This has been verified for all slit orientations independently of the input parameters. 

Consequently, we use \begin{math} \delta < 0.5 \times R_\text{star}/P_s \end{math} as a rule-of-thumb for the maximal displacement of the deepest thin lines. 
We use this relation as an upper limit for whether the displacement can be observed, i.e. when \begin{math} \delta \ge 0.01 \end{math} pixel. 
We consider here that a displacement below one percent of a pixel may no longer be detectable owing to pixel inhomogeneities in the detector. 
As a result, we can predict those configurations where it should not be possible to retrieve the position angle, as seen in Table \ref{tab:prediction}, and compare these predictions to the following simulation results. 

The simulation also gives us a means to study the accuracy and the precision of the extraction. We consider a good detection to be when the retrieved position angle is within $10^\circ$ of the input value. Measures with errors over $15^\circ$ are considered as poor.

\renewcommand{\arraystretch}{1.3}
\begin{table}[h]
\caption{Predictions for the signal detection}
\label{tab:prediction}
\centering
\begin{tabular}{l| c c c c} \hline \hline 
\cline{1-5}
  &  \multicolumn{4}{c}{Plate scale} \\
$R_\text{star}$ & 0.1"/pix & 0.15"/pix & 0.2"/pix & 0.5"/pix \\
\hline 
2.5 mas &  N.V. & N.V. & N.V. & N.V. \\
5 mas & 2 \% & 1.3 \% & N.V. & N.V. \\
10 mas & 4 \% & 2.6 \% & 2 \% & N.V. \\
15 mas & 6 \% & 4 \% & 3 \% & 1.2 \% \\ \hline
\end{tabular}
\tablefoot{Signal detection predictions, giving the photocentre displacement in percentage of a pixel. Any displacement below 1 \% is accounted as not visible (N.V.)}
\end{table}

\subsection{Direct extraction}

The direct extraction method only requires the position spectrum for the analysis. 
It relies on measuring the photocentre displacement directly from the position spectrum. 
When the spectro-astrometric signal is at least twice stronger than the continuum, which occurs when the  ratio stellar radius over pixel plate scale is maximized, this method delivers position angle estimates up to  2$^\circ$ accuracy.
The relevant simulation results are shown in Table \ref{tab:directh}.
On the plate scale and in stellar radius range where, according to the previous predictions, the determination of the position angle is possible, the measure yields an accuracy of $2^\circ$. 
In addition, for a field of view of 1"/pix for a stellar radius of 2.5 mas, the extracted value still gives a reasonable order of magnitude with an error of about $10^\circ$.

\renewcommand{\arraystretch}{1.3}
\begin{table}[h]
\caption{Retrieved position angles using direct extraction}
\label{tab:directh}
\centering
\begin{tabular}{l|c c c}

\hline \hline
 & \multicolumn{3}{c}{$\text{PA}_\text{measured} - \text{PA}_\text{true}$ (in deg)} \\ 
  $R_{star}$ &  0.086 "/pix & 0.1"/pix & 0.15 "/pix  \\\hline 
2.5 mas & 7.6 $\pm$ 10.9 & 10.3 $\pm$ 10.9 &  -- \\
5 mas  & 0.6 $\pm$ 3.7  & 2.4 $\pm$ 7.0 & -- \\
10 mas &  2.2 $\pm$ 1.8  & 1.5 $\pm$ 3.9 & -0.7 $\pm$ 17.2 \\
15 mas  &  -2.9 $\pm$ 1.2 & -2.8 $\pm$ 2.9 & 2.11 $\pm$ 14.7 \\
\hline
\end{tabular}
\tablefoot{Only the relevant results are given here.
Configurations with higher plate scale deliver errors above $30^\circ$ for the position angle.}
\end{table}

\subsection{Cross-correlation extraction}

Unlike the direct extraction, the cross-correlation extraction makes use of both the intensity and the position spectrum. Indeed, the cross-correlation function is calculated between the position spectrum and the derivative of the intensity spectrum.

\begin{figure*}
\begin{subfigure}[t]{0.49\textwidth}
\includegraphics[width = \textwidth]{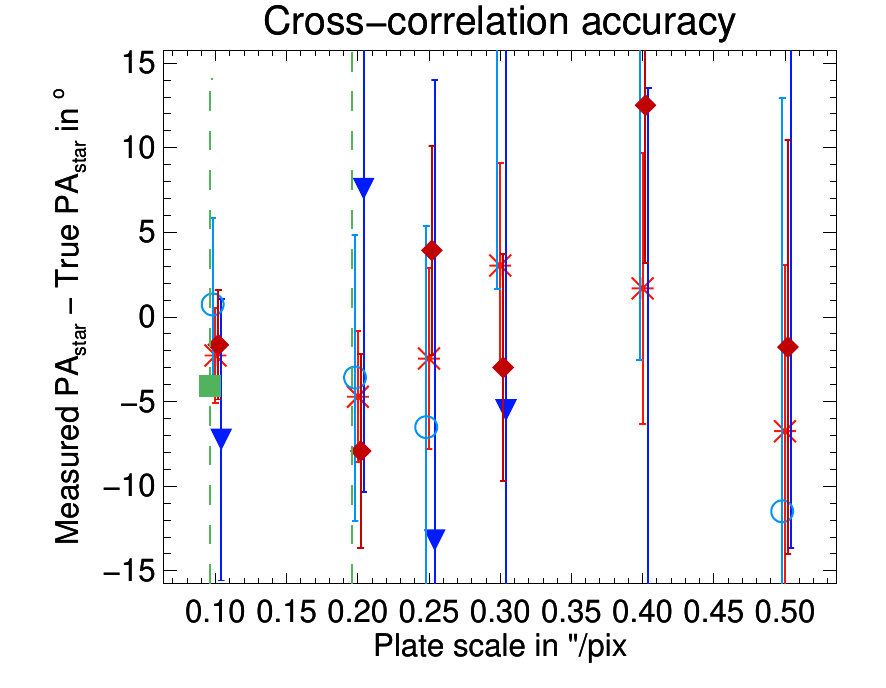} 
\end{subfigure} \hspace{1em}
\begin{subfigure}[t]{0.49\textwidth}
\includegraphics[width = \textwidth]{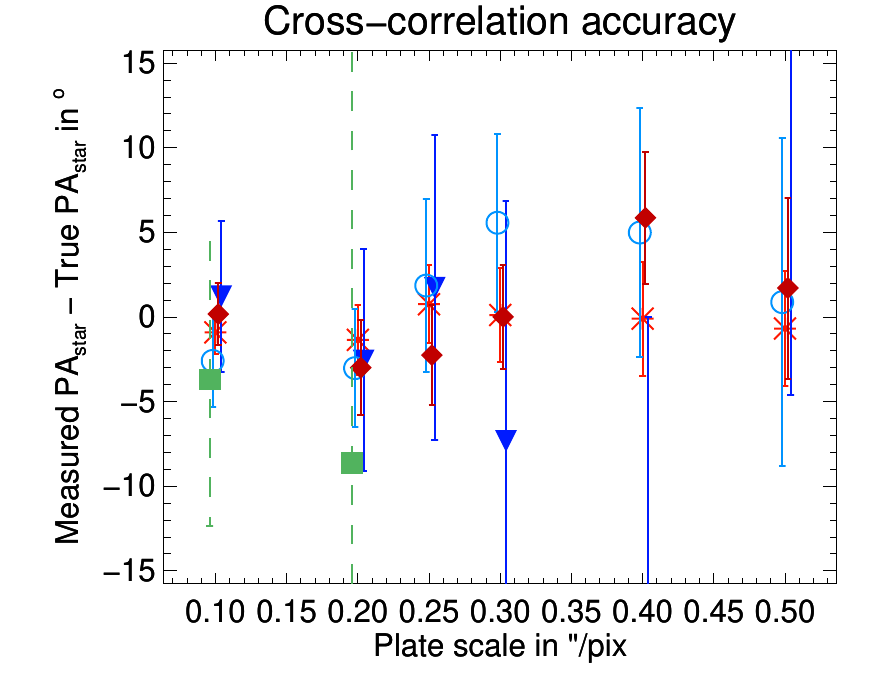}
\end{subfigure}
\caption{Left panel: accuracy of the cross-correlation method using the errors of a single cross-correlation function. Right panel: accuracy of the refined method using the standard deviation between several cross-correlation functions for identical slit orientation as error estimate. The absolute stellar position angle is estimate for a high resolution spectra (R=120\,000) and for increasing apparent stellar radii. Filled squares: $R_\text{star}$= 1 mas; filled triangles: $R_\text{star}$= 2.5 mas; circles: $R_\text{star}$ = 5 mas; filled diamonds: $R_\text{star}$= 10 mas and stars: $R_\text{star}$= 15 mas. To improve the visibility of the error bars, the points for a given plate scale are slightly dispersed around that value.}
\label{fig:resulthigh}
\end{figure*}

The mathematical evaluation of the cross-correlation errors has been discussed by \citet{Zucker} who proposes using a likelihood estimator. 
This approach to estimating the errors was studied in our simulations. 
However, it yields extremely high errors for the cross-correlation function when slit and spin axis are parallel, where it is expected that the cross-correlation function remains close to nought owing to the absence of signal in the position spectrum. 
We have tried to modify the relation of \citet{Zucker} for our case, in order to limit the error amplitude when the cross-correlation function is close to zero, but such that it is still sensitive to slight fluctuations when the cross-correlation function is close to one, by using the formula 
\begin{math} \sigma = (1-C_{xy})^2 / 2W \end{math}. 

The cross-correlation method delivers slightly better results than the direct method, especially when the signal does not domi\-nate the position spectrum. 
As illustrated in Fig \ref{fig:resulthigh} for small plate scales and large stellar diameters, i.e. when the signal is potentially directly measurable, then the accuracy of the determined position angle is comparable to what is achieved using the direct extraction method. 
In addition, when the ratio \begin{math}R_\text{star}/P_s \end{math} approaches or is less than 1\%, the retrieved value of the position angle is dominated by more than $10^\circ$ errors. 
However, while any plate scale above 0.15"/pix leads to very imprecise and inaccurate angles in the direct extraction method, this threshold is moved towards larger plate scales with the cross-correlation method. 
For instance, for a stellar radius of 15 mas and a plate scale of 0.40"/pixel leading to a \begin{math} R_\text{star}/P_s\end{math} ratio of 3.75\%, the retrieved position angle is $2^\circ \pm 7^\circ$ which is still within the $10^\circ$ error range.   

The simulations produced similar results for the high resolution case R=60\,000 with identical rotational velocity, so these are not reproduced here. It implies that as long as the spectrograph resolves the rotational broadened stellar lines, it is not necessary to go for higher resolution. As a result the choice of the resolution of the spectrograph is set by the rotational velocity of the target star.

\subsection{Refinement of the cross-correlation analysis}

In practice, at high signal-to-noise, the cross-correlation errors are highly dependent on the noise in both the position and intensity spectrum. They are thus caused mostly by photon noise.
 Therefore, several spectra at identical orientation $\psi$ and similar signal to noise ratio should deliver identical cross-correlation functions. 
This property was incorporated in the model by creating a set of ten noise-independent simulated 2D spectra per orientation. 
For each 2D spectrum from a set, slight variations in the seeing were allowed, and the noise properties were calculated anew to reproduce real observing conditions as much as possible. The spectra are reduced as before, extracting both position and intensity spectrum. 
Each matching pair of position and intensity spectra are then cross-correlated as previously performed. Finally the errors are calculated as the standard deviation of the different cross-correlation functions. 

This analysis method requires a specific observing strategy where, for each orientation, a set of  spectra is recorded. 
Instead of co-adding the spectra, each is reduced independently to gain statistical information on the noise of the global measure. 
In the case of a Coud\'e spectrograph, where the uncorrected slit orientation varies with the hour angle, the number of consecutive exposures may be limited in order to keep a relatively constant value of $\psi \pm 5^\circ$ per set. 

By using this method, the large error bars observed previously in Fig~\ref{fig:resulthigh} are lowered and the accuracy of the measure is improved. This is directly related with the choice of the errors estimate: while for small plate scale we expect a strong signal at orthogonal slit-spin orientations, this signal decreases for increasing plate scales. As a result the maxima of the cross-correlation function may no longer be close to one. 
Figure \ref{fig:resulthigh} illustrates the new determined position angles using ten spectra at identical orientation to build the errors of the cross-correlation functions per orientation. The signal-to-noise for each individual spectrum is constant for a given orientation. 

The refined method uses the noise properties of a set of spectra to evaluate the cross-correlation errors. 
As a result, the accuracy of the retrieved measurement of the position angle is improved, and the corresponding errors are divided by a factor two compared to the unrefined method. 
Only when the signal is already very strong does it not improve the measurement made by the direct method. 
In addition, the detection limit estimated earlier of 1\% of a pixel displacement is also reached. Indeed, as illustrated in Fig~\ref{fig:resulthigh}, the position angle retrieved for the 5 mas radius star, at the largest plate scale  is still within our acceptable error range.  
This emphasizes how the choice of the instrument is relevant for a given target in order to optimize the validity of the measure of the absolute stellar position angle. 
Finally, using the refined error evaluation method, even stars with 1 mas radius could be observed with fewer than $10^\circ$ errors in the measured position angle. 

\subsection{Signal-to-noise}

The previous results were obtained with a raw S/N in the 2D spectra around 150. We made the assumption that the spectra are dominated by photon noise. In that case, the raw S/N is estimated as the square root of the mean peak value of the order. 
However, the real S/N of the data is determined from the cross-correlation function. 
Using the method described in the previous section to calculate the errors of the cross-correlation, the true S/N of our data is then
\begin{math} S/N = \bar{p} / \delta_p \end{math}, where $\bar{p}$ is the average peak value of the cross-correlation function, and $\delta_p$ the error calculated for this point. 

To emphasize the relation between raw S/N and true S/N, we let the raw S/N vary for one stellar radius at one fixed spectrograph plate scale, and measured both the true S/N from the cross-correlation function, and the extracted position angle. For each given raw S/N, the simulation was run 100 times to assess the detection robustness. 
The results are summarized in Table \ref{tab:SNR}. To reach a $3\sigma$ detection, a raw S/N of 100 is necessary. However the scatter in the retrieved position angles remains large, with more than $10^\circ$. 
Nevertheless, this also shows that detections above the $5\sigma$ threshold are also possible without resorting to spectra co-addition.

\begin{table}
\caption{Scatter in the retrieved position angles}
\label{tab:SNR}
\centering
\begin{tabular}{c c c} \hline \hline
Raw S/N & True S/N & $3\sigma$ scatter \\ 
100 & 3.0 & 11.35 \\
134 & 4.4 & 7.74 \\
163 & 5.0 & 5.17 \\
190 & 5.8 & 4.47 \\
256 & 8.3 & 3.76 \\ \hline
\end{tabular}
\tablefoot{The $3\sigma$ scatter is calculated from the distribution of the retrieved position angles for increasing signal. }
\end{table}

\subsection{Potential targets}

The refined cross-correlation method opens the possibility of determining the stellar position angle for a large sample of stars. The potential target should satisfy three constraints. First it has to have a large number of deep spectral lines for the cross-correlation method to deliver the best measurements. This condition is fulfilled by G-type and cooler stars. Second, the apparent diameter of the star should be big enough -- from our simulation results at least 2 mas. Finally, the stellar lines should remain as narrow as possible. 
The last constraint points towards slowly rotating stars, but also towards low turbulences in the stellar atmosphere. 

\begin{figure}
\resizebox{\hsize}{!}{\includegraphics{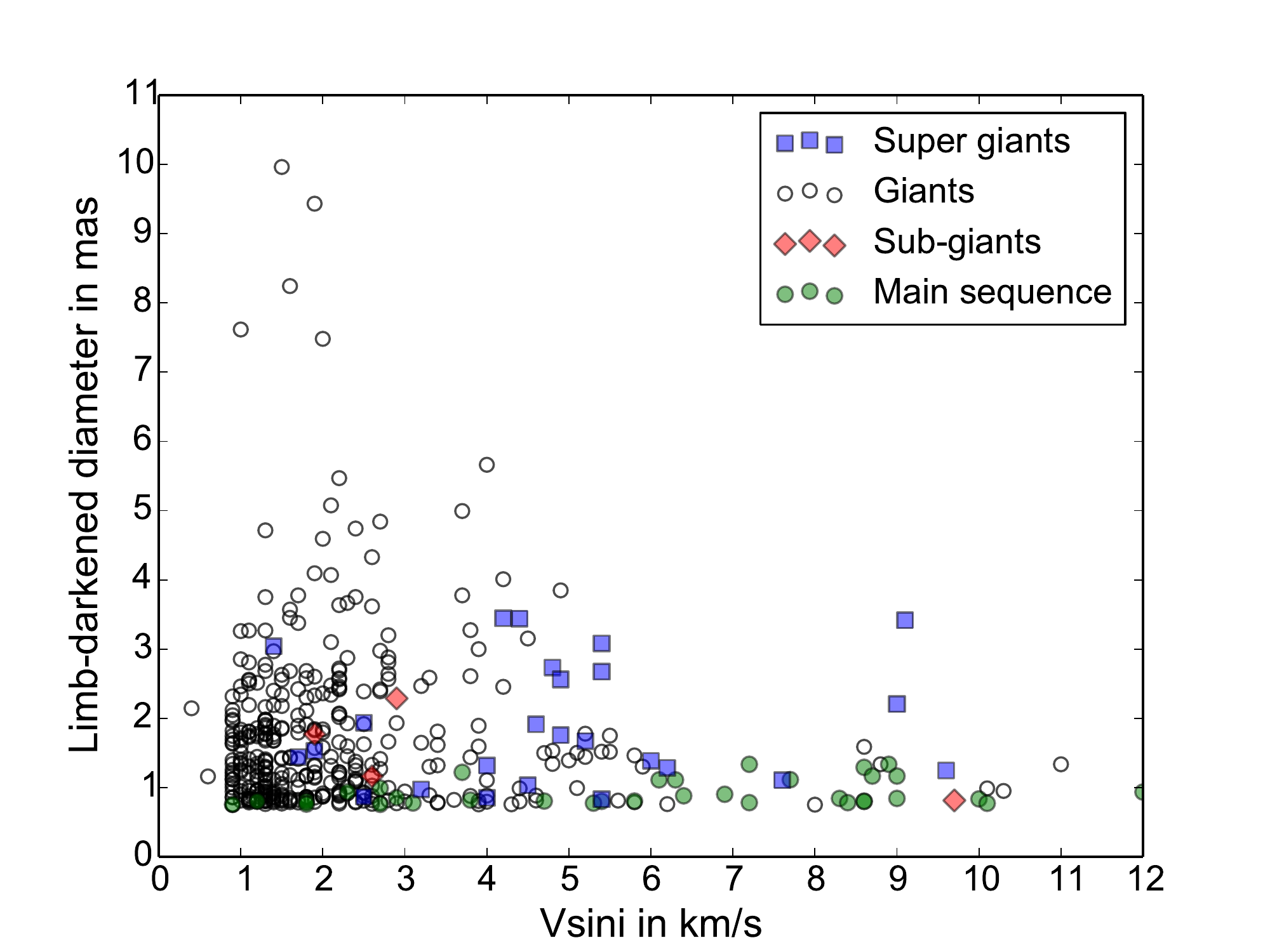}}
\caption{Properties of the potential target candidates for the spin axis determination. The sample is limited here to stars with apparent diameters larger than 0.75 mas. Mostly of the potential candidates are cool K-giants. A large majority of the main sequence stars are slowly rotating A-type stars. }
\label{fig:stellardist}
\end{figure}

Figure \ref{fig:stellardist} illustrates the spectral class distribution of the potential targets. Most targets are K giants. However, a few sub-giants and super-giants could also be observed. The main-sequence stars are identified as slowly rotating A-type stars. 
Consequently, the potential targets span a large sample of evolution states at the end of a star's life.

\begin{figure*}[th]
\hspace{1em}
\begin{subfigure}[t]{0.425 \textwidth}
\centering
\includegraphics[width = \textwidth]{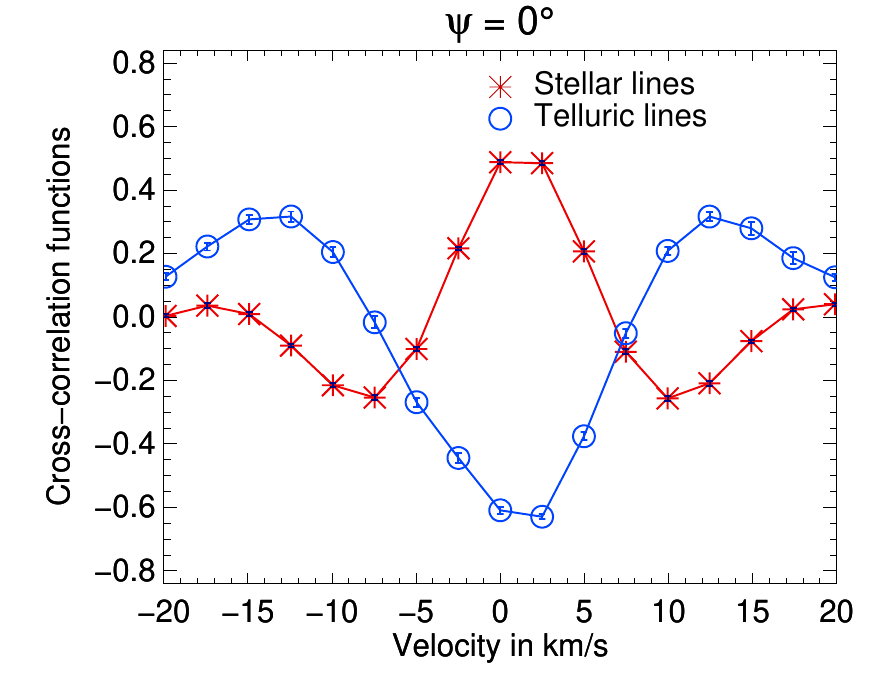}
\end{subfigure}
\hspace{4em}
\begin{subfigure}[t]{0.425\textwidth}
\centering
\includegraphics[width = \textwidth]{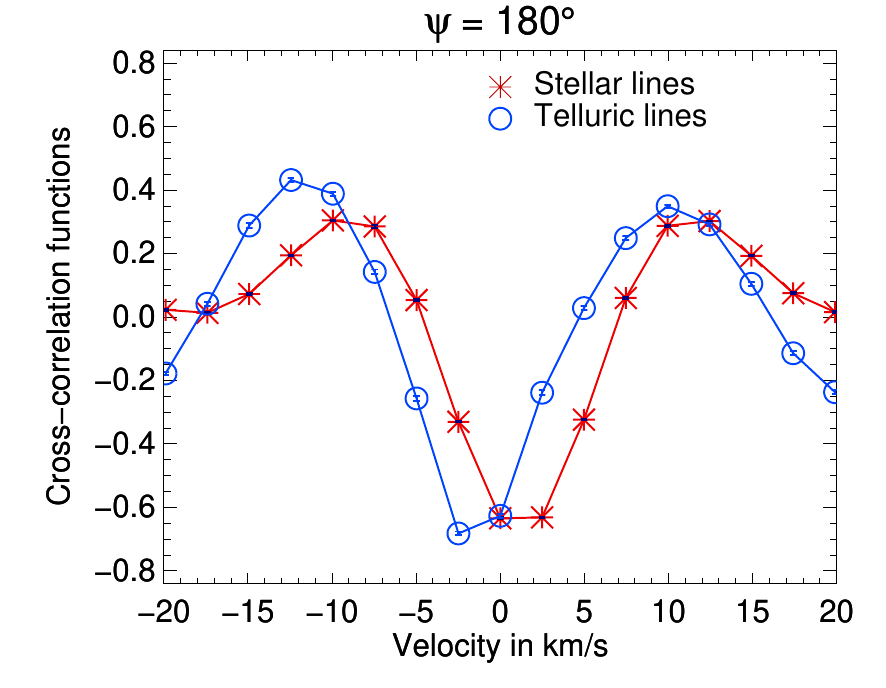}
\end{subfigure}
\caption{Disentangling the instrumental and the stellar contribution to the line tilt. Left panel: the stellar rotation induces a line tilt in the opposite direction to the instrumental tilt. Right panel: the stellar rotation induces a line tilt in the same direction as the instrumental tilt. The simulated stellar position angle is $PA_\text{star} = 75^\circ$. Maximum displacement is expected for $\phi = -15^\circ \text{ and } 195^\circ$. Each cross-correlation function is calculated by selecting only telluric and stellar lines respectively. The instrumental contribution remained constant between the two slit orientations. }
\label{fig:seingtilt}
\end{figure*}

\section{Discussion}\label{sec:discuss}

The simulations performed previously assumed, for simplicity, optimal observing conditions and a basic stellar model. The  results are now compared to real observations: non-optimal observing conditions and more realistic stellar properties. We first investigated the impact of seeing and instrumental variations on the results. 

\subsection{Impact of seeing and instrumental profile}

The shape and the size of the seeing point spread function (hereafter seeing PSF) are likely to fluctuate during an observing night. 
Owing to the short time-scale for seeing variations, the seeing FWHM can change drastically between two consecutive exposures. 
\cite{Brannigan} did an intensive study of the influence of seeing and slit width on the position spectra. 
We summarize the main ideas here. If the seeing FWHM is smaller or on scale with the slit width, it causes artefacts in the spectrum. 
These can have a shape similar to the stellar rotation signature once it has been reduced. In order to limit and to remove these artefacts, they recommend using a very narrow slit. 
This condition has already been implemented in our model by stating that the slit width is five times smaller than the seeing FWHM. 
In addition, they advise the observers to take spectra at anti-parallel slit orientations, e.g. $0^\circ$ and $180^\circ$.  
Indeed, artefacts are in theory independent of the slit orientation and remain constant while the stellar signal is inverted at $180^\circ$ slit orientation. 
Artefacts are thus removed from the position spectrum by subtracting anti-parallel orientations while the stellar signal adds up.

Moreover, the instrumental profile is also likely to add further distortions in the spectral lines. 
For instance, slight astigmatism or misalignment in the camera optics can lead to a small tilt, a bending, or a displacement of the lines along the spatial direction. 
It is then necessary to distinguish the target signal from the one caused by the instrument. 
If the slit width is very narrow compared to the mean seeing FWHM, the instrument profile is the main contributor to the lines distortion. 
With a careful analysis of the spectra, it is, however, possible to disentangle the stellar rotation signal from the instrumental artefacts. 
We take as example the case where the instrumental profile induces a line tilt comparable in amplitude to the stellar rotation signal. 
The instrumental tilt is then present on all the lines of the spectrum. It superimposes the stellar tilt on all but the telluric lines by using the latter as the reference for the state of the instrumental profile, the stellar tilt is retrieved. As illustrated in the Fig~\ref{fig:seingtilt}, the stellar lines do not have the same cross-correlation functions as the telluric lines, despite an instrumental tilt.  
In addition, since the instrumental profile varies on a long time scale (several tens of minutes if the spectrograph is directly attached to the telescope, to a few hours if it is stored in a confined room) it is approximately constant between two consecutive slit orientations. 
As a result, variations in the stellar line tilt can still be monitored.

In practice, it may not always be possible to have a very narrow slit width compared to the seeing FWHM. 
In such cases, the artefacts induced by seeing on the spectra can no longer be ignored, and it is a necessity to take spectra at anti-parallel slit orientations to remove those artefacts. 
The short time scale of the seeing fluctuation means that it is not possible to take two consecutive images with exactly identical seeing conditions, which compromises the anti-parallel subtraction. 
As a result, we launched the development and use of an additional optic, a differential image rotator, whose purpose is to project two images of the star next to each other at different  orientations on the slit (\cite{lesage}). The two resulting spectra would then be taken with identical seeing conditions and are suitable to anti-parallel subtraction.

\subsection{Limb-darkening}

\begin{figure*}
\centering
\includegraphics[width = 18cm]{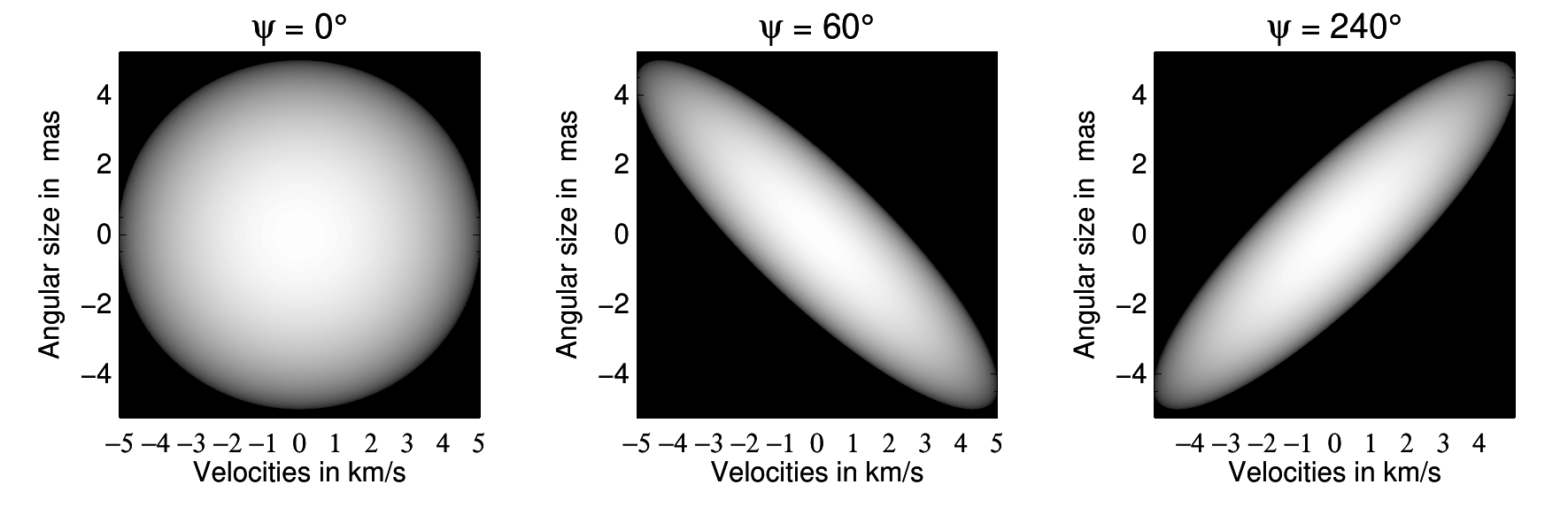}
\caption{Doppler profile in the case of differential rotation for three values of $\psi$ = $0^\circ$, $60^\circ$ and $240^\circ$. The rotation rate at the equator is set at 5 km/s, and the relative differential parameter $\alpha = 0.2$.}
\label{fig:diffdop}
\end{figure*}

We used a linear limb-darkening law for our model that produces a satisfactory approximation,  considering the precision of our measure. 
The wavelength dependence of the limb darkening coefficient was not implemented in the model. 
As the apparent stellar limb increases with wavelength, so does the spectro-astrometric signal. On average, using the measurement of \cite{LaFrasse}, the apparent diameter increases around 6\% from the B to the K band.

\subsection{Stellar spots}

Stellar spots have already been mapped using spectro-astrometry on TW Oph, a C5.5 star, and RS Vir, a M6 star, by \cite{Voigt} using the \textsc{Crires} instrument. 
In their method, the stellar rotation was a source of error that had to be removed before analysis. 
They included the line tilt caused by the rotation in the instrumental PSF and applied PSF decorrelation. 
As a result, only the signal of the spot remained. 
In our case, the spot is a perturbation source, whose impact has to be ascertained.

Firstly, if the spot and the star have identical absorption spectra, differing only by the spot temperature, then the continuum of the position spectrum is homogeneously shifted towards the spot, when it is a hot spot, and away from the spot if it is a cool spot. 
However, spot and star usually do not have identical spectra since the cool spot allows the observer to see more deeply into the star's interior and be affected by different elements than at the surface. 
There are two possibilities for the photocentre to be shifted significantly: either when the star has an absorption line while the spot has none or when the spot is absorbing at a wavelength where the star is not. The spot itself has no influence on the tilt of the line. 

We now assume a single circular spot of intensity $I_{spot}(\lambda)$ on the stellar surface. The stellar intensity is noted $I_{*}(\lambda)$. 
The radius of the spot is characterized by \begin{math} a = R_\text{spot}/R_\text{star} \end{math}. The position of the spot along the slit direction is defined by \begin{math} b = y_\text{spot} / R_\text{star} \end{math}. In a case without limb darkening and using these two parameters and a simple barycentric relation, the displacement $B(\lambda)$ of the photocentre is calculated via
\begin{equation}
B(\lambda) = \frac{R_{star}\; b\; I_{spot}(\lambda) \; a^2}{a^2 I_{spot}(\lambda) + I_{*}(\lambda) \;(1-a^2)}.
\end{equation}

According to this equation, maximum displacement is expected when $I_{*}(\lambda)$ tends towards 0, $B(\lambda)$ being proportional to $y_spot$ the position of the spot on the star. The latter cannot be greater than the stellar radius. 
The spectro-astrometric signature presents a unique shift in the photocentre towards the spot at the wavelength of the stellar absorption line.
The displacement amplitude remains smaller than the rotational signal because the former is proportional to the spot position on the stellar surface.

\subsection{Differential rotation}

According to predictions (\cite{Kitchatinov}), K giants are susceptible to presenting differential rotation in their atmosphere. 
Recent measurements by \cite{Weber} have shown that the relative differential parameter $\alpha$, which defines the rotation gradient from the equator velocity $\Omega_o$ to the pole as a function of the solar latitude $l$
\begin{equation}\label{eq:diffrot} \Omega (l) = \Omega_o (1 - \alpha \sin^2 l) \end{equation},
is $\alpha \le 0.05$ for K0 to K2 giants. 
Relation (\ref{eq:diffrot}) was implemented in Equation (\ref{eq:dwdop}) by setting \begin{math} Y = R_{star} \sin l \end{math}. 

\begin{figure}
\resizebox{\hsize}{!}{\includegraphics{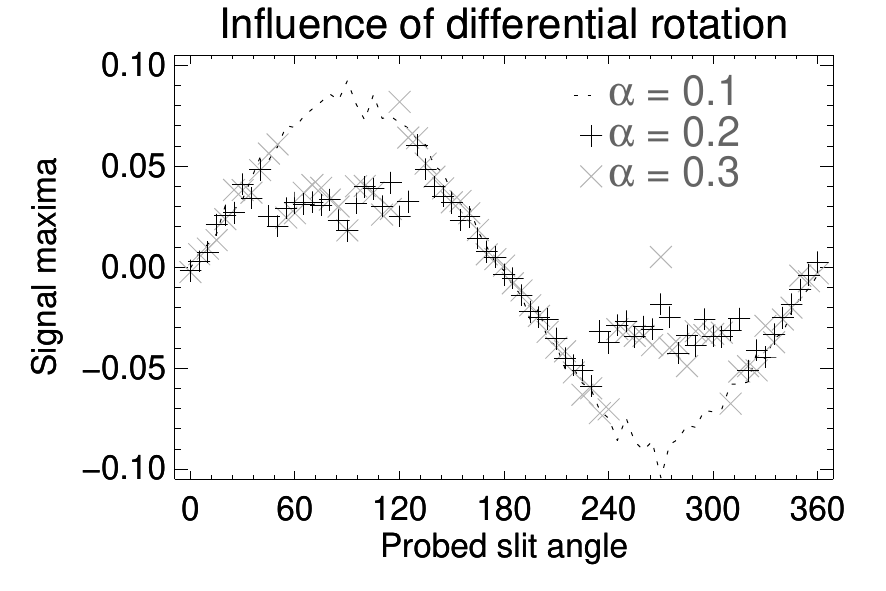}}
\caption{Evolution of the signal amplitude with increasing differential rotation parameter. The signal maxima are extracted directly from the position spectrum, using an optimized configuration. }
\label{fig:diffalpha}
\end{figure}

However, since some of our potential target stars are F and G giant stars, we allowed $\alpha$ to vary between 0 and 0.4. 
Strong differential rotation visibly affects the shape of the Doppler rotation profile, as seen in Fig \ref{fig:diffdop}. 
The faster rotating equator deforms the elliptical profile towards an asymmetrical profile. 
This effect is then reflected in the position spectrum at the slit angle supposed to maximize the spectro-astrometric signal. 
This effect is reproduced in Fig~\ref{fig:diffalpha}. For low values of $\alpha$, the retrieved displacement maxima still follow a sine curve, but for increasing $\alpha$ values, the larger displacements are clipped.

The presence of differential rotation in the star affects the profile of the amplitude variations by cutting the extreme of the sine curve. We suspect that it remains possible to determine the position angle using the remaining portion of the profile for $\alpha$ up to 0.3 by dramatically increasing the sampling of $\psi$. 
In addition, there are several existing methods, e.g. the Fourier transform method used by \cite{Reiners} on cool stars, to look for differential rotation in stars that could be applied beforehand to ensure that the observed target presents minimal or moderate differential rotation. 
Were the star to present important differential rotation, the observing strategy would then include additional slit position in order to sample sufficiently the general profile, and adding differential rotation to the model fit.


\section{Conclusion}\label{sec:conclu}

We have shown that stellar rotation induces a sub-pixel tilt in the stellar spectral lines whose inclination depends on the absolute stellar position angle. 
By using a spectro-astrometric analysis of high resolution spectra, we proved that it is possible to determine the absolute stellar position angle with $10^\circ$ accuracy for a moderate sample of stars. 
Our simulations demonstrated two possible ways to retrieve the spectro-astrometric signal, i.e. direct measurement of the photocentre displacement or cross-correlation analysis.
The appropriate method depends on the signal-to-noise of the spectra, the observing set-up, i.e. the spectrograph's plate scale, the target, i.e. the apparent stellar diameter, and the slit width. 
However, we have shown that there is a optimal combination between plate scale and stellar radius that optimizes the detection rates for a constant raw signal-to-noise ratio. 

According to our results, with an instrument like CRIRES, which provides a plate scale of 0.086"/pix when coupled with an adaptive optic, it is possible to determine the position angle of stars with apparent diameter down to 2 mas with errors that are lower than $10^\circ$. 
The observations should be taken for at least four different slit angles, and can be obtained either sequentially with seeing effects, or simultaneously using additional instrumentation.
This represents a sample of about 100 targets, from A-type main sequence to various K-type giants, including a handful of known spectroscopic binaries. 
Probing the stellar position angle for these targets would provide a first step in answering two of the original questions: can the stellar spin axis change with the stellar evolution, and how well are the spin axes of binaries aligned.

\begin{acknowledgements}
A-L. L gratefully acknowledge the support from the German Federal Ministry for Research and Technology (BMBF grant 05A08BU2) and from Hamburg University. This study made use of Rene Heller's Holt-Rossiter-McLaughlin Encyclopedia.
\end{acknowledgements}

\bibliographystyle{aa}
\bibliography{desspot_aa}

\end{document}